\documentclass[10pt,conference]{IEEEtran}

\usepackage{mypreamble}
\begin{document}

\title{Human-In-the-Loop Software Development Agents}

\makeatletter
\newcommand{\newlineauthors}{%
  \end{@IEEEauthorhalign}\hfill\mbox{}\par
  \mbox{}\hfill\begin{@IEEEauthorhalign}
}
\makeatother

\author{\IEEEauthorblockN{Wannita Takerngsaksiri\IEEEauthorrefmark{1}, Jirat Pasuksmit\IEEEauthorrefmark{2}, Patanamon Thongtanunam\IEEEauthorrefmark{3}, Chakkrit Tantithamthavorn\IEEEauthorrefmark{1},\\ 
Ruixiong Zhang\IEEEauthorrefmark{2},
Fan Jiang\IEEEauthorrefmark{2}, 
Jing Li\IEEEauthorrefmark{2},
Evan Cook\IEEEauthorrefmark{2}, 
Kun Chen\IEEEauthorrefmark{2}, 
Ming Wu\IEEEauthorrefmark{2}}
\IEEEauthorblockA{\IEEEauthorrefmark{1}Monash University, Australia.}
\IEEEauthorblockA{\IEEEauthorrefmark{3}The University of Melbourne, Australia.}
\IEEEauthorblockA{\IEEEauthorrefmark{2}Atlassian, Australia.}
}


\maketitle
\thispagestyle{plain}
\pagestyle{plain}

\begin{abstract}

Recently, Large Language Models (LLMs)-based multi-agent paradigms for software engineering are introduced to automatically resolve software development tasks (e.g., from a given issue to source code). 
However, existing work is evaluated based on historical benchmark datasets, rarely considers human feedback at each stage of the automated software development process, and has not been deployed in practice.
In this paper, we introduce a Human-in-the-loop LLM-based Agents framework (HULA) for software development that allows software engineers to refine and guide LLMs when generating coding plans and source code for a given task. 
We design, implement, and deploy the HULA framework into Atlassian JIRA for internal uses.
Through a multi-stage evaluation of the HULA framework, Atlassian software engineers perceive that HULA can minimize the overall development time and effort, especially in initiating a coding plan and writing code for straightforward tasks.
On the other hand, challenges around code quality remain a concern in some cases.
We draw lessons learned and discuss opportunities for future work, which will pave the way for the advancement of LLM-based agents in software development.
\end{abstract}





\section{Introduction}

Multi-agent paradigm, powered by Large Language Models (LLMs), has demonstrated remarkable results in many research areas~\cite{zhao2023survey}, including software engineering~\cite{liu2024large, hou2023large, zan2022large}.
Within the software engineering context, an LLM-based multi-agent is defined as an autonomous agent (e.g., an LLM model) receiving inputs from the surrounding environment (e.g., compilers, linter tools) and performing actions (e.g., writing code, fixing bugs) to achieve a specific objective (e.g., resolving a pull request, passing test cases).
The agent will be assigned with a role (e.g., software engineer, software tester) as part of the software development process with access to external resources (e.g., compilers, linters).
With this LLM-based multi-agent paradigm, they are able to solve complex software development tasks for a given input and a goal.


Recently, various LLM-based software development agent frameworks are proposed to address complex software development tasks~\cite{yang2024swe, zhang2024autocoderover, chen2024coder, tao2024magis, xia2024agentless, arora2024masai, ma2024understand}.
However, existing LLM-based software development work is (1) evaluated based on historical benchmark datasets of open-source software projects, (2) does not consider human feedback at each stage of the automated software development process, and (3) has not been deployed in practice.
These challenges raise important questions of whether the effectiveness of LLM-based software development agents can be generalised beyond the context of open-source software projects, how acceptable the plans and code generated by LLM-based software development agents at each stage of the software development process, and how practitioners perceive the benefits and challenges of using LLM-based software development agent.

\emph{In this paper}, we introduce a \underline{Hu}man-in-the-\underline{L}oop LLM-based Software Development \underline{A}gent Framework (\ours).
Our framework consists of three agents: AI Planner Agent, AI Coding Agent, and Human Agent, working cooperatively to achieve the common goal of resolving a JIRA issue (i.e., a software development task).
We design the user interface, implement, and deploy \ours{} into Atlassian JIRA for internal uses.
To evaluate its effectiveness, our evaluation consists of three stages: (1) an offline evaluation of \ours{} without human feedback using SWE-Bench and our internal dataset of 369 JIRA issues; (2) an online evaluation of \ours{} augmented by human feedback using real-world 663 JIRA issues; and (3) an investigation of the practitioners' perceptions on the benefits and challenges of using \ours{} in practice.
Through the multi-stage evaluation of our \ours{} through real-world deployment and user studies among Atlassian practitioners, we answer the following three research questions:

\begin{enumerate}[{\bf RQ1)}]
\item {\bf (Offline) \rqone}\\
For SWE-bench, AI Planner Agent can correctly identify files that need to be changed, achieving an average recall of 86\% per issue, while AI Coding Agent can correctly generate code that is similar to the human-written code (i.e., ground-truth) with an average similarity score of 45\%.
On the other hand, when applying Agents to the Atlassian internal dataset, AI Planner Agent achieves an average recall of 30\% per issue, while AI Coding Agent achieves an average similarity score of 30\%.
We observed that our JIRA issues tend to be shorter than the open-source software projects.
\item {\bf (Online) \rqtwo}\\
Based on the online evaluation of \ours{}, plans are successfully generated for 527 out of the 663 real-world JIRA issues.
Of these, practitioners approved the generated plans for 433 out of 527 plan-generated issues, achieving a plan approval rate of 82\%. 
During the coding stage, pull requests (PRs) were created for 95 out of 376 code-generated issues, achieving a raised PR rate of 25\%.
Of these, 56 PRs were successfully merged, leading to a merged PR rate of 59\%.
These results highlight the success of the \ours{} framework where human feedback is incorporated into practice. 

\item {\bf (Survey) \rqthree}\\
41\% of participants agreed that the generated plan by \ours{} aligned accurately with the issue description.
While \ours{} still requires human involvement to ensure the generated code completely addresses the task,  61\% of the participants agreed that the generated code was easy to read and modify, which helped reduce their initial development time and effort.
Additionally, participants acknowledged that \ours{} promotes good documentation practice, though this requires more effort to provide detailed issue descriptions.



\end{enumerate}



\textbf{Contributions:} To the best of our knowledge, this paper is the first to present: 
\begin{itemize}
    \item The human-in-the-loop LLM-based software development agents framework (\ours) integrated into Atlassian JIRA (Section~\ref{sec:autodev}).
    \item The multi-stage evaluation of LLM-based software development agents, focusing on offline and online evaluation, together with practitioners' survey among Atlassian practitioners (Section~\ref{sec:evaluation}).  
    \item The findings, lessons learned, and future directions from the multi-stage evaluation and deployment of \ours{} in practice (Section~\ref{sec:findings}). 
\end{itemize}


\textbf{Paper Organization.}
The paper is organized as follows.
Section~\ref{sec:background} describes the software development workflow at Atlassian.
Section~\ref{sec:motivation} discusses related works and limitations.
Section~\ref{sec:autodev} presents \ours~framework. 
Section~\ref{sec:evaluation} presents the multi-stage evaluation framework.
Section~\ref{sec:findings} presents our results.
Section~\ref{sec:future} discusses the lessons learned and future directions.
Section~\ref{sec:threat} discloses the threats to validity.
Finally, Section~\ref{sec:conclusion} draws the conclusion.

\section{Software Development Workflow at Atlassian}\label{sec:background}

Atlassian is an enterprise software company that develops products for software development, project management, team collaboration, and many more.
Atlassian offers tools such as JIRA Software (for task management), Confluence (for collaboration and knowledge management), and Bitbucket (for source code management) to enhance software teams' productivity, streamline workflows, and facilitate Agile methodologies across various industries.
These products 
serve over 300,000 customers globally, spanning various industries such as technology, finance, healthcare, and government. 

At Atlassian, practitioners use JIRA\footnote{https://www.atlassian.com/software/jira} as a central tool for managing software development tasks, tracking bugs, and facilitating collaboration across teams.
First, they will create JIRA issues to define features, tasks, or bug fixes to work on.
Then, JIRA issues will be assigned to individual practitioners or teams.
Once they start working on the given JIRA issue, practitioners will create a coding plan (e.g., which files need to be modified or fixed?, where to fix a bug?), aiming to identify files that are relevant to the given JIRA issue (i.e., code or bug localization).
Then, they will start the coding process.
Once the coding process is finished, they will raise a pull request for code review and automated testing, ensuring that each pull request is of sufficient quality before integrating it into the main repository.

In 2024, Atlassian had over 12,000 engineers submitting over 950,000 pull requests.
Since these processes are generally done manually, the teams are prone to productivity challenges, e.g., an overwhelming backlog of JIRA issues and context switching between tools.
These challenges can result in delays and inefficiencies in software delivery, prompting Atlassian teams to seek automated solutions that can streamline workflows and improve their overall productivity.

\section{Related Works and Limitations}\label{sec:motivation}




Automated software development tools and techniques, powered by deep learning (DL) or large language models (LLMs), are proposed to assist software engineers in various software development tasks (e.g., bug localization~\cite{Zhou2012bugs,tantithamthavorn2013using,tantithamthavorn2018impact}, code completion~\cite{takerngsaksiri2024syntax, takerngsaksiri2024student}, code generation~\cite{arora2024masai, tao2024magis}, test case generation~\cite{ takerngsaksiri2024tdd,alagarsamy2024a3test}, code review automation~\cite{li2022codereviewer, pornprasit2024gpt,thongtanunam2022autotransform,hong2022commentfinder}, vulnerability detection and repair~\cite{fu2022linevul,li2024llm,fu2024aibughunter,fu2022vulrepair}).
However, these automated software development tools and techniques are mainly designed to tackle individual SE tasks, leading to suboptimal performance when applied to real-world software development tasks.

Recently, an LLM-based multi-agent paradigm for software engineering has emerged~\cite{liu2024large}. 
The multi-agent paradigm for software engineering involves multiple autonomous agents, each with specific roles (e.g., software engineer, testers) and goals (e.g., writing code, writing test cases), working together to solve complex problems. 
Various LLM-based software development agent frameworks are proposed to resolve complex software development tasks~\cite{yang2024swe, zhang2024autocoderover, chen2024coder, tao2024magis, xia2024agentless, arora2024masai, ma2024understand}.
For example, Yang~\ea~\cite{yang2024swe} proposed SWE-agent, an LLM-based agent built on top of a Linux shell, enabling the agent to effectively search, navigate, edit, and execute code commands.
Zhang~\ea~\cite{zhang2024autocoderover} proposed AutoCodeRover, an LLM-based agent with code search on AST (i.e., Abstract Syntax Tree) and spectrum-based fault localization using a test suite.
Ma~\ea~\cite{ma2024understand} proposed RepoUnderstander, an LLM-based agent with a Monte Carlo tree search-based exploration strategy on the repository knowledge graph.
Tao~\ea~\cite{tao2024magis} proposed Magis, a multi-agent framework with four customized roles: manager, repository custodian, developer, and quality assurance engineer.
Chen~\ea~\cite{chen2024coder} proposed CodeR, a multi-agent framework with pre-defined task graphs.
Arora~\ea~\cite{arora2024masai} proposed Masai, a modular framework that divided the problem into multiple sub-problems for LLM-based agents.

\textbf{Limitations.} However, existing LLM-based software development work is (1) evaluated based on historical benchmark datasets, limiting the understanding of its effectiveness in the context of resolving enterprise software development tasks; (2) rarely considering human feedback at each stage of the automated software development process, limiting the understanding of the practitioners' acceptability of the plans and code generated by LLM-based software development agent at each stage to resolve enterprise software development tasks; and (3) has not been deployed in practice, limiting the understanding of practitioners' perceptions on the benefits and challenges of using LLM-based software development agent.

\begin{figure*}[t]
    \centering
    \includegraphics[width=\textwidth]{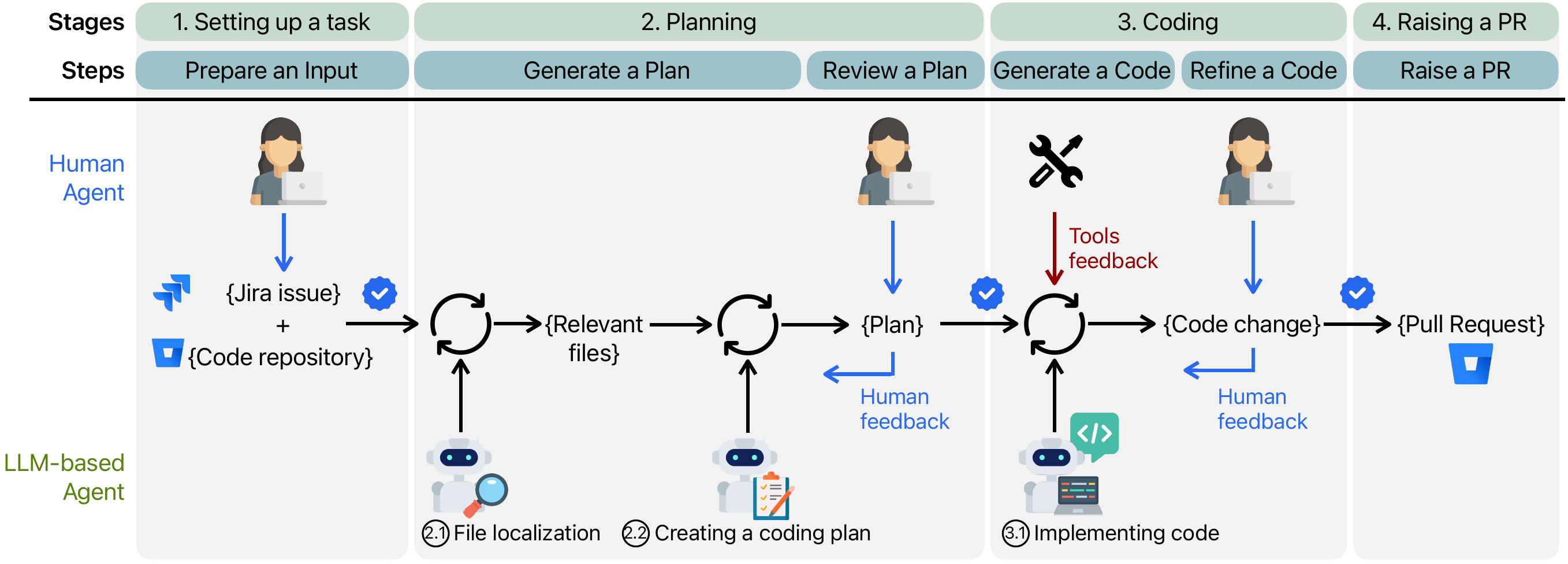}
    \caption{An Overview of our \underline{Hu}man-in-the-\underline{L}oop LLM-based Software Development \underline{A}gents Framework (\ours), consisting of three agents (i.e., AI Planner Agent, AI Coding Agent, and Human Agent) work cooperatively to achieve the common goal of resolving a JIRA issue.}
    \label{fig:architecture}
\end{figure*}

\section{\ours: Human-in-the-loop LLM-based Agents Framework for Software Development}\label{sec:autodev}

In this section, we present a human-in-the-loop LLM-based agent framework for software development.

At Atlassian, human intelligence (i.e., the expertise of practitioners) and human authority (i.e., the human ability to rule AI agents) are highly valued. 
Therefore, rather than aiming to fully automate software development tasks, we designed an LLM-based software development agent to collaborate with practitioners, functioning as an assistant to help resolve software development tasks, promoting the Human-AI synergy~\cite{lo2023synergy,Hoda2023augmented}.
We envision that in the future of software development, practitioners and LLM-based software development agents should work collaboratively to complete software development tasks.
Thus, we design our \ours{} framework to incorporate human feedback at each step of the software development tasks.
This allows practitioners to provide feedback and guidance as needed, ensuring the software development process remains collaborative rather than entirely reliant on fully autonomous agents.

Below, we present the overview of our \ours{} framework (see Figure~\ref{fig:architecture}) and the user interface of \ours{} being seamlessly integrated into Atlassian JIRA (see Figure~\ref{fig:autodev-workflow}).

\begin{figure*}[t]
    \centering
    \includegraphics[width=\textwidth]{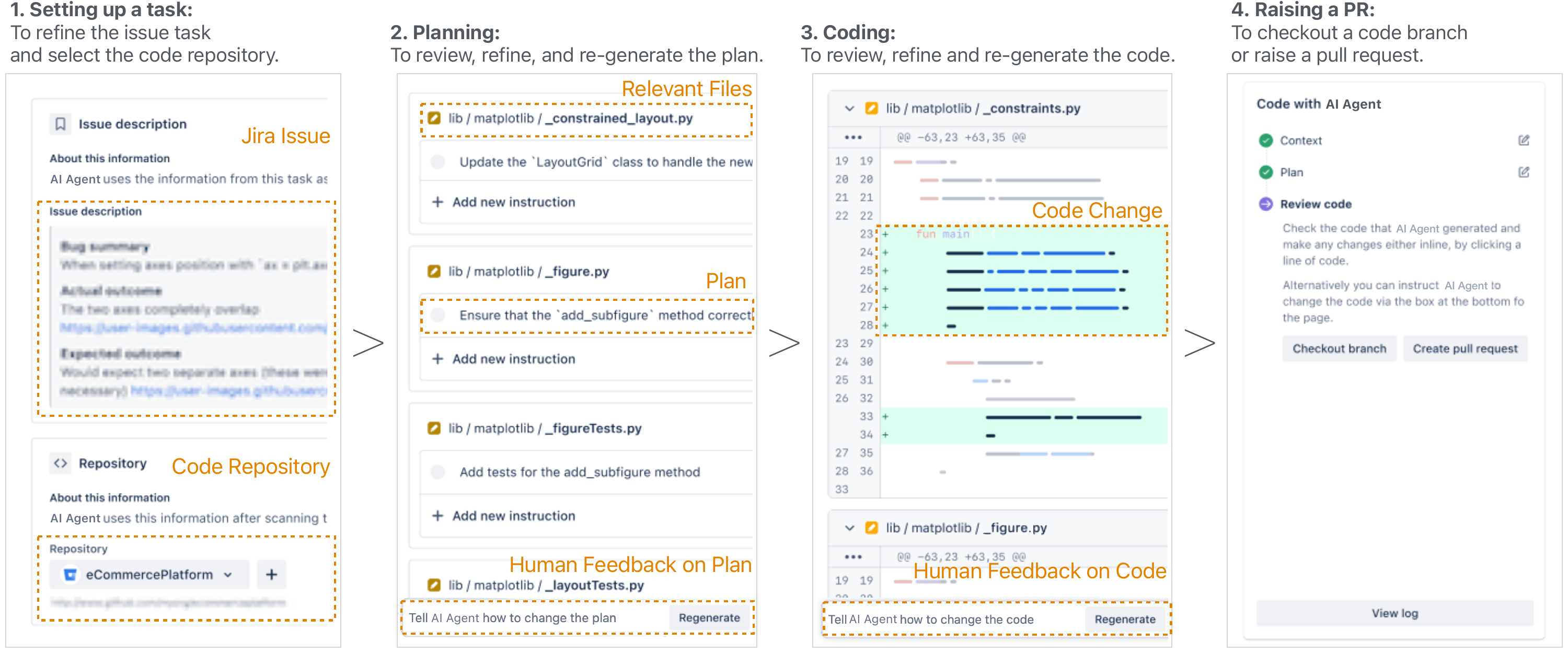}
    \caption{The user interface of our human-in-the-loop LLM-based agent framework, seamlessly integrated into Atlassian JIRA.}
    \label{fig:autodev-workflow}
\end{figure*}

\subsection{Framework}

The ultimate goal of our human-in-the-loop LLM-based software development agent framework is to generate a pull request (i.e., a set of code changes) based on a given software development task (i.e., a JIRA issue).
Since software development tasks for enterprise systems are complex, we exploit a multi-stage method to dissect the software development task into distinct stages, namely, \emph{file localization} (i.e., which files should be changed?), \emph{plan generation} (i.e., how each file should be changed), and \emph{coding} (i.e, what are the suggested code changes?).
Formally speaking, our framework consists of the following three agents:

\begin{itemize}
    


    \item AI Planner Agent ($\mathcal{P}$) is an LLM model receiving inputs (i.e., a JIRA issue and the source code repository) to identify the more relevant files that are associated with the given JIRA issue (i.e., file localization ~\cite{tantithamthavorn2013using, tantithamthavorn2018impact,zhou2012should}). 
    Then, the LLM model uses the list of relevant files and the JIRA issue to generate a coding plan (e.g., ``add tests for the add\_subfigure method").
    
    \item AI Coding Agent ($\mathcal{C}$) is an LLM model receiving inputs (i.e., the coding plan) to generate a set of code changes.
    
    \item Human Agent ($\mathcal{H}$) is a software engineer assigned to the given task. The human agent will provide feedback and work cooperatively with AI agents, particularly, reviewing and editing the plan and the code, and raising PRs, to achieve the common goal (i.e., resolve a JIRA issue).
\end{itemize}

With the Human-AI synergy in mind, we leverage a human-AI agent coordination paradigm, exploiting human intelligence in each stage to enhance the effectiveness of the multi-agent system for software development and facilitate better alignment with human preference.
Different from the other multi-agent systems that leverage competitive, mixed, or hierarchical coordination styles without human involvement, our framework exploits the cooperative communication style with the Decentralized Planning Decentralized Execution (DPDE) paradigm to ensure that each AI agent is independently responsible for its own objective and capabilities while being able to work cooperatively with human agents to achieve the common goal.
Although we exploit the DPDE paradigm (i.e., no communication among agents), all Agents still have access to the shared memory of information (i.e., the JIRA issue and the source code repository).
The merit of this approach lies in minimal communication overhead, reducing the computational resources on the central LLM, and enhancing the Agents' adaptability, as each agent can quickly adjust its behaviour based on local information and human feedback. 
This makes our framework better suited for dynamic and unpredictable environments while promoting human-AI synergy.



Our framework consists of the following four stages.  


\textbf{Stage \circled{1} Setting up a task:} A software engineer begins the workflow by selecting a development task (i.e., a JIRA issue) and an associated code repository.
Generally, the issue description contains necessary documented information for humans to understand the work to be done, e.g., a user story, definition of done, acceptance criteria, or code examples~\cite{pasuksmit2021towards}.

\textbf{Stage \circled{2} Planning:} 
AI Planner Agent ($\mathcal{P}$) uses information in the JIRA issue (i.e., the issue summary and description) to understand the work to be done and the context and identify files that are relevant to the given JIRA issue. 
The software engineer can review, edit, and confirm the information before proceeding to the next step.
Once the relevant files are determined, the AI Planner Agent generates a coding plan, outlining how the code in the identified files should be changed to resolve the given JIRA issue.
For example, ``Add tests for the add\_subfigure method".
After the coding plan is generated, the Human Agent ($\mathcal{H}$) can review the plan, provide additional instructions, and re-generate the coding plan.
Alternatively, the Human Agent ($\mathcal{H}$) can directly modify the list of relevant files, and edit the change plan for each file.


\textbf{Stage \circled{3} Coding:}
Once the Human Agent ($\mathcal{H}$) approves the coding plan, the LLM-based agent proceeds to the execution of the given coding plan.
The AI Coding Agent ($\mathcal{C}$) generates the code changes to each file according to the plan. 
The Human Agent ($\mathcal{H}$) can then review the proposed code changes.
If the code change does not meet the expectation, the Human Agent ($\mathcal{H}$) can provide further instructions to change the code and then ask the AI Coding Agent ($\mathcal{C}$) to re-generate the code changes.
To ensure that the code being generated is syntactically correct with high code quality~\cite{takerngsaksiri2024syntax}, the AI Coding Agent ($\mathcal{C}$) undertakes self-refinement, improving the generated code based on feedback from code validation tools such as compilers and linters.
The refinement is conducted iteratively until the code successfully passes the validation tools or the maximum number of attempts is reached.

\textbf{Stage \circled{4} Raising a Pull Request:} 
Once the Human Agent ($\mathcal{H}$) agrees with the code change, the generated code changes are created as a pull request (PR) to BitBucket for review by other practitioners.
Alternatively, the Human Agent ($\mathcal{H}$) can create a new code branch from the generated code to make modifications prior to the PR creation.

\section{The Multi-Stage Evaluation of \ours{} in Real-World Deployment}\label{sec:evaluation}

In this section, we introduce the multi-stage evaluation framework for evaluating our \ours{} in real-world deployment.
Our evaluation aims to answer the following three research questions.
\begin{itemize}
    \item RQ1: \rqone
    \item RQ2: \rqtwo
    \item RQ3: \rqthree
\end{itemize}

To answer these research questions, our evaluation consists of three stages, i.e., offline evaluation, online evaluation, and practitioner survey.
The main purpose of offline evaluation is to validate the performance of \ours{} prior to the actual deployment.
Since the offline evaluation did not incorporate human feedback and cannot capture the real-world usage, we conducted an online evaluation to investigate how practitioners use \ours{} in practice.
However, the online evaluation still lacks the user perception.
To address this, we run a practitioner survey to investigate the performance of \ours{} from user perspectives.
Below, we present the research methodology for each stage of the evaluation (see Figure~\ref{fig:eval-overview}).

\begin{figure}
    \centering
    \includegraphics[width=\linewidth]{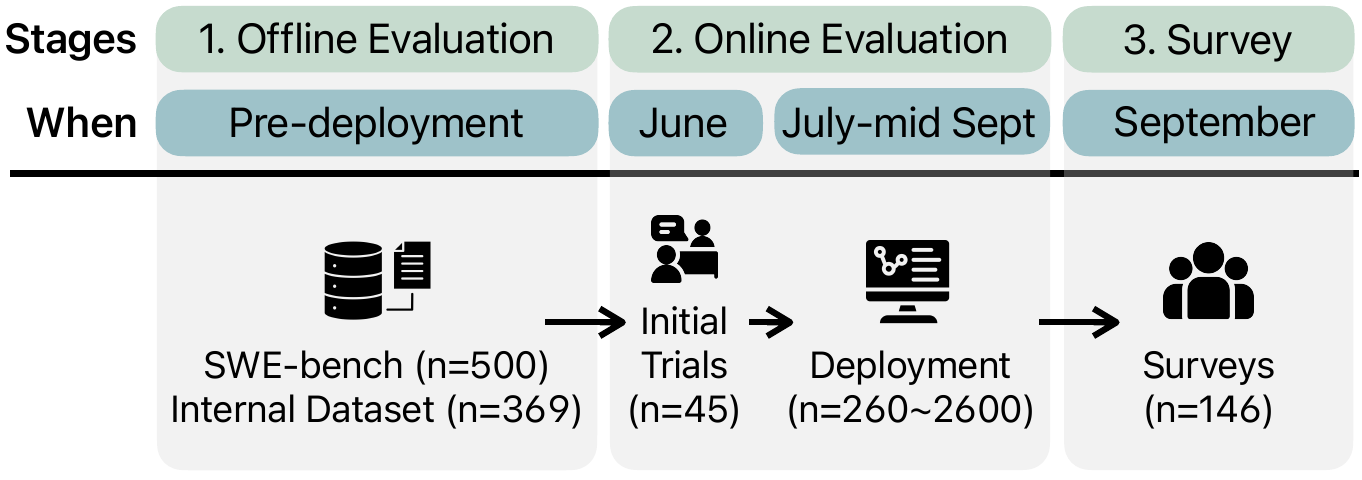}
    \caption{An Overview of our Multi-stage Evaluation of  \ours{}.}
    \label{fig:eval-overview}
\end{figure}

\subsection{Stage 1: Offline Evaluation}
The offline evaluation based on benchmark datasets allows us to preliminary assess the performance of \ours{} to ensure that it achieves an acceptable performance before deployment.
Note that we did not include human feedback for this offline experiment to fully automate the evaluation process.   

\textbf{Benchmark Datasets:}
To conduct an offline evaluation, we need high-quality benchmark datasets that provide sufficient information of the work to be done and human-written code solutions for diverse programming languages.  
Thus, we use the SWE-bench dataset which provides high-quality datasets for Python programming language.
Then, we collected development issues at Atlassian to evaluate the effectiveness of \ours{} in an industrial setting that also covers diverse contexts and programming languages.
Table \ref{tab:stat} provides an overview statistics of our benchmark datasets.

\begin{itemize}
    \item \textit{SWE-bench Verified~\cite{sweverify2024}} is a subset of SWE-bench~\cite{jimenez2023swe} that has been verified for appropriately scoped unit tests and well-specified issue descriptions by human annotators.
    The dataset consists of 500 development tasks (i.e., Github issues), corresponding code solutions, and unit tests written by developers. 
    The dataset is collected from 12 open-source Python repositories on GitHub, including various types of development tasks like bug reports, code refactoring suggestions, and feature requests.

    \item \textit{Internal dataset} is a collection of completed development tasks (i.e., JIRA issues) at Atlassian.
    We randomly sampled issues from \textit{JIRA Software} with the corresponding pull requests (PRs) from BitBucket.
    In total, our internal dataset contains 369 JIRA issues from 94 software repositories.
    The issues' context ranges from front-end developments, back-end developments, data analytics, security patches, product migrations, configuration updates, bug fixes, and more.
    The code solutions in PRs involve various file types with more than 10 programming languages including TypeScript, Java, Kotlin, and Python.
\end{itemize}

\textbf{Metrics:} 
To evaluate the performance of \ours{} based on the benchmark datasets, we focus on the following four aspects: success generation rate, file localization, functionality correctness, and the similarity of the generated code and the ground truth source code (i.e., human-written code changes).
To do so, we input each JIRA issue into our \ours{} framework and measure the generated plans and code based on the following five metrics:

\begin{table}[t]
    \caption{A Statistics Summary of Benchmark Datasets.}
    \centering
    \resizebox{\linewidth}{!} {%
    \begin{tabular}{l|l|c|c|c}
        \textbf{Dataset} & \textbf{Data} & \textbf{Min} & \textbf{Median} & \textbf{Max} \\
        \hline
        \multirow{3}{*}{\makecell[l]{SWE-bench \\ Verified \\ (n=500)}} 
        & Issue Information (token) & 24 & 295 & 6,939 \\
        & Changed files (count) & 2 & 2 & 22\\
        & Human-written code (token) & 89 & 220 & 5,057\\
        
        \hline
        \multirow{3}{*}{\makecell[l]{Internal \\(n=369)}} & Issue 
        Information (token) & 11 & 75 & 1,114\\
        & Changed files (count) & 1 & 3 & 44\\
        & Human-written code (token) & 82 & 1,275  & 47,520 \\
    \end{tabular}
    }
    \label{tab:stat}
\end{table}

\begin{itemize}
    \item \textit{\%Issues for Success Generation} measures the percentage of issues that \ours{} can successfully generate plans and code, completing all of the steps in the workflow (see Figure~\ref{fig:architecture}) (i.e., success generation).
Otherwise, if \ours{} fails at any step in the workflow and is unable to produce the final code changes, we consider it as unsuccessful. 

    \item \textit{Recall of File Localization} measures the percentage of correct files that need to be changed for a given issue.

    \item \textit{\%Issues for Perfect File Localization} measures the percentage of issues that achieve a perfect file localization (i.e., all files recommended by AI Planner Agent are exactly the same as the actual files that need to be changed---no false positives and no false negatives).


    \item \textit{\%Issues for Perfect Passing Test Cases} measures the percentage of issues that achieve 100\% of passing all corresponding unit test cases (i.e., functionality correct).
    Due to the unavailability of unit testing in our internal dataset, this measure is only applicable to the \textit{SWE-bench Verified} dataset only, not our internal dataset.
    


    \item \textit{\%Issues of High Code Similarity} measures the percentage of issues that the AI-generated code is highly similar to the ground-truth human-written code. 
    Code similarity is used as a proxy to determine if the code being generated is highly similar to the ground-truth human-written code, due to the absence of the test cases in our internal dataset.
    Typically, human involvement is typically required to determine if the code is similar or not, yet it is very expensive in practice.
    Recently, CodeBERTScore~\cite{zhou2023codebertscore} or BERTScore~\cite{zhang2019bertscore} are widely used in software engineering to evaluate code similarity.
    However, these techniques are designed for BERT-like architectures, limiting its input length to only 512 tokens, which does not cover the typical size of the human-written code (i.e., a median of 1,275 tokens for our internal dataset).
    Following the LLM-as-a-Judge principle\cite{zheng2023judging}, we exploit GPT-4 as a judge to measure the similarity between the source code and the ground-truth, given the corresponding JIRA issue as a context.
    After iterations of prompt refinement, we develop a prompt instruction containing \emph{$<$prompt, issue, generated code, ground-truth code$>$} to generate the similarity score ranging from 1 (no similarity), 2 (weak similarity), 3 (high similarity), to 4 (highly similarity).
    Any issue that achieves a code similarity score of 3 to 4 is considered high code similarity.
\end{itemize}



\subsection{Stage 2: Online Evaluation}
We conducted an online evaluation to assess the performance of \ours{} in the actual development practice with practitioners at Atlassian.
This evaluation allows us to gain further insights from actual usage conditions.

\textbf{Internal Deployment:}
To avoid interruption in the day-to-day work of Atlassian practitioners, we opted to deploy this new technology carefully as per the following steps.
First, we conducted initial trials with a small group of software engineers from two software development teams to gain initial feedback.
After ensuring positive feedback from the first group, we gradually deploy \ours{} with a broader group of practitioners.
Figure \ref{fig:autodev-workflow} shows an example of \ours{} workflow that integrated into a JIRA issue.

\textit{Initial Trials.} 
We deployed \ours {} to a small group of software engineers (called an alpha group).
This alpha group comprises 45 engineers across two software development teams working on different products.
We made \ours{} available to the alpha group for two weeks and monitored the usage, performance, and reliability.
We also collected feedback from the alpha group to gain deeper insights into user experiences and the quality of the generated code.
The feedback of the alpha groups is mainly related to the user interfaces and experience of \ours{}. 
We conducted several iterations of trials to stabilize the design of our \ours~workflow.

\textit{Deployment.}
After improving \ours{} based on initial trials, we expanded the deployment to broader Atlassian teams. 
To minimize impacts to large teams, we implemented a gradual, incremental deployment. 
We began rolling out \ours{} to practitioners in early July 2024, starting with 260 practitioners. By mid-September 2024, the deployment expanded to 2,600 practitioners.
In total, \ours{} were made available to more than 22,000 eligible issues.


\textbf{Metrics:}
In this online evaluation, we observe the effectiveness of \ours{} using the real-world user interaction measured by the following metrics:

\begin{itemize}
    \item \textit{Plan Generation Rate} measures the percentage of issues that \ours{} can generate a coding plan, compared to the total number of issues with attempted plan generation.

    \item \textit{Plan Approval Rate} measures the percentage of issues that practitioners approve the generated coding plan of  \ours{}, compared to the total number of issues with successful plan generation.
    
    \item \textit{Code Generation Rate} measures the percentage of issues that \ours{} can complete the workflow and generate code changes, compared to the total number of issues with attempted code generation.

    \item \textit{Raised PR Rate} measures the percentage of issues that practitioners clicked to raise a pull-request (PRs) based on \ours{}-generated code, compared to the total number of issues with successful code generation.

    \item \textit{Merged PR Rate} measures the percentage of issues with PRs that practitioners select to merge the PRs into the repository's main branch, compared to the total number of issues with a PR raised.
    


\end{itemize}

\subsection{Stage 3: Practitioners' Survey}

To gather deeper insights on \ours{} quality and feedback, we conducted an online survey with practitioners who have used \ours{} to generate code (i.e., reach the \textit{refine code} step) at least once. 
The survey is designed to capture the performance of \ours~in the internal deployment and to gather the user perceptions (e.g., benefits, challenges, and areas of improvement).
We conduct the pilot survey with five software engineers to validate and refine the questions prior to the full launch.
In the full launch, we sent a survey invitation directly to the practitioners via Slack communication.

\textit{Survey.} The survey began with demographic questions to collect background information of participants on their job titles, experiences in IT, experiences with AI coding assistants, and the frequency of usage on \ours{}.
Next, the participants were asked to rate their level of satisfaction with the accuracy and effectiveness of \ours{} in the planning and coding stages. 
Finally, they were asked to provide their perceptions through open-ended questions.
Excluding the demographic questions, the survey consists of 11 questions (see Table~\ref{tab:survey_questions}).

\textit{Analysis.} To analyze the open-ended responses, the first and the second authors conducted an open coding~\cite{charmaz2014constructing} analysis in an online session to extract corresponding themes of each response.
This was followed by three rounds of card sorting~\cite{spencer2009card} with the predefined themes to systematically categorize the responses into categories. 
We assessed the inter-rater reliability by calculating Cohen's Kappa coefficient~\cite{mchugh2012interrater}. 
The mean score of Kappa values across open-ended questions is 0.67,
reflecting a substantial level of agreement among the evaluators.

    
    

\begin{table}[t]
    \caption{Survey questions (excluding demographics questions).}
    \label{tab:survey_questions}
    \resizebox{\linewidth}{!} { %
    \begin{tabular}{p{0.05\linewidth}|p{0.9\linewidth}}
        \textbf{Item} & \textbf{Question} \\
        \hline
        & \textbf{The performance of \ours{}} \\
         & Please rate your satisfaction with the following aspects of Autodev. \\
        Q1$^\ddagger$ & The identified files are relevant to the issue description. \\
        Q2$^\ddagger$ & The identified files align with how I would have approached this issue. \\
        Q3$^\ddagger$ & The changes to the identified files align accurately with the issue description. \\
        Q4$^\ddagger$ & The changes to the identified files align accurately with how I would have approached this issue. \\
        Q5$^\ddagger$ & The generated code can be easily understood and modified to solve the issue. \\
        Q6$^\ddagger$ & The generated code accurately solves the issue.\\
        Q7$^\ddagger$ & The generated code contains no defects and fulfils non-functional requirements. \\
        Q8$^\ddagger$ & The generated code change is complete and fully addresses the issue. \\
        \hline
        & \textbf{Feedback for \ours{}} \\
        Q9* & What are the main benefits you received from using Autodev? \\
        Q10* & What are the main challenges you experienced from using Autodev? \\
        Q11* & How can we improve Autodev for you? \\
        \hline
    \end{tabular}
    }
    \scriptsize{$^\ddagger$ Likert scale of agreement: $\bigcirc$ Strongly agree, $\bigcirc$ Somewhat agree, $\bigcirc$ Neutral,\\ $\bigcirc$ Somewhat disagree, $\bigcirc$ Strongly disagree}
    
    \scriptsize{* Optional open-ended questions.}
\end{table}

\section{Results}\label{sec:findings}
In this section, we present the results according to our research questions.
        

\subsection*{\textbf{RQ1: \rqone}}

Table~\ref{tab:offline} shows the evaluation results based on SWE-bench and the internal datasets.
Specifically, Table~\ref{tab:offline} shows that \ours{} can complete the generation workflow for 97\% of the issues in the SWE-bench dataset and 100\% of the issues in our internal dataset.
This indicates that \ours~has high success in the generation.
We now discuss the effectiveness of \ours{} in each benchmark dataset.

Table~\ref{tab:offline} shows that \ours{} can correctly identify all the files (i.e., Perfect File Localization) for 84\% of the issues in SWE-bench.
The code generated by \ours{} achieved a similarity score of 45\% compared to human-written code and passed the unit tests for 31\% of the issues. This performance is comparable to SWE-agent Claude, which ranks 6th on the leaderboard as of 26 September 2024.
These results indicate that \ours{} performs reasonably well on SWE-bench.





Table~\ref{tab:offline} shows that \ours{} achieved a perfect file localization on 15\% of the issues and a high similarity score compared to human-written code on 30\% of the issues in our internal dataset.
The lower performance could be due to various factors.
One possible factor is the diversity of the data. The SWE-bench dataset is limited to the Python programming language, whereas our internal dataset includes more than 10 programming languages and over 20 file types. 
Moreover, the SWE-bench dataset was collected from 12 open-source software repositories, while our internal dataset was sourced from 94 software repositories. 
This increased diversity in programming languages and repositories may challenge the LLM agent's generalizability, leading to suboptimal performance.

Another potential factor is the context information and complexity of the changes. The issues in the SWE-bench dataset are more detailed, with a median length of 295 tokens, compared to 75 tokens in our dataset. 
Furthermore, the SWE-bench dataset has fewer changed files with a median of two files per issue, while our internal dataset has a median of three files per issue. 
These differences make the task more challenging to identify the correct files for changes, leading to a decrease in file localization accuracy in our internal dataset.

\begin{tcolorbox}[boxsep=0mm, left=2mm, right=2mm]
\textbf{Finding 1.} 
For SWE-bench, AI Planner Agent can correctly identify files that need to be changed, achieving an average recall of 86\%, while AI Coding Agent can correctly generate code that is similar to the human-written code with an average similarity score of 45\%.
On the other hand, for our internal dataset, AI Planner Agent achieves an average recall of 30\%, while AI Coding Agent achieves an average similarity score of 30\%.
\end{tcolorbox}


\begin{table}[t]
    \centering
    \caption{(RQ1) The offline evaluation results of \ours{}.}
    \resizebox{\linewidth}{!}{
    \begin{tabular}{l|c|c}
         \textbf{Metrics} & \textbf{SWE-bench Verified} & \textbf{Internal} \\
         \hline
         \%Issues for Success Generation & 97\% & 100\% \\ 
         Recall of File Localization & 86\% & 30\% \\
         \%Issues for Perfect File Localization & 84\% & 15\% \\
         \%Issues for Perfect Passing Test Cases & 31\% & - \\ 
         \%Issues of High Code Similarity & 45\% & 30\% \\
    \end{tabular}
    }
    \label{tab:offline}
\end{table}

\begin{figure}[t]
    \centering
    \includegraphics[width=\columnwidth]{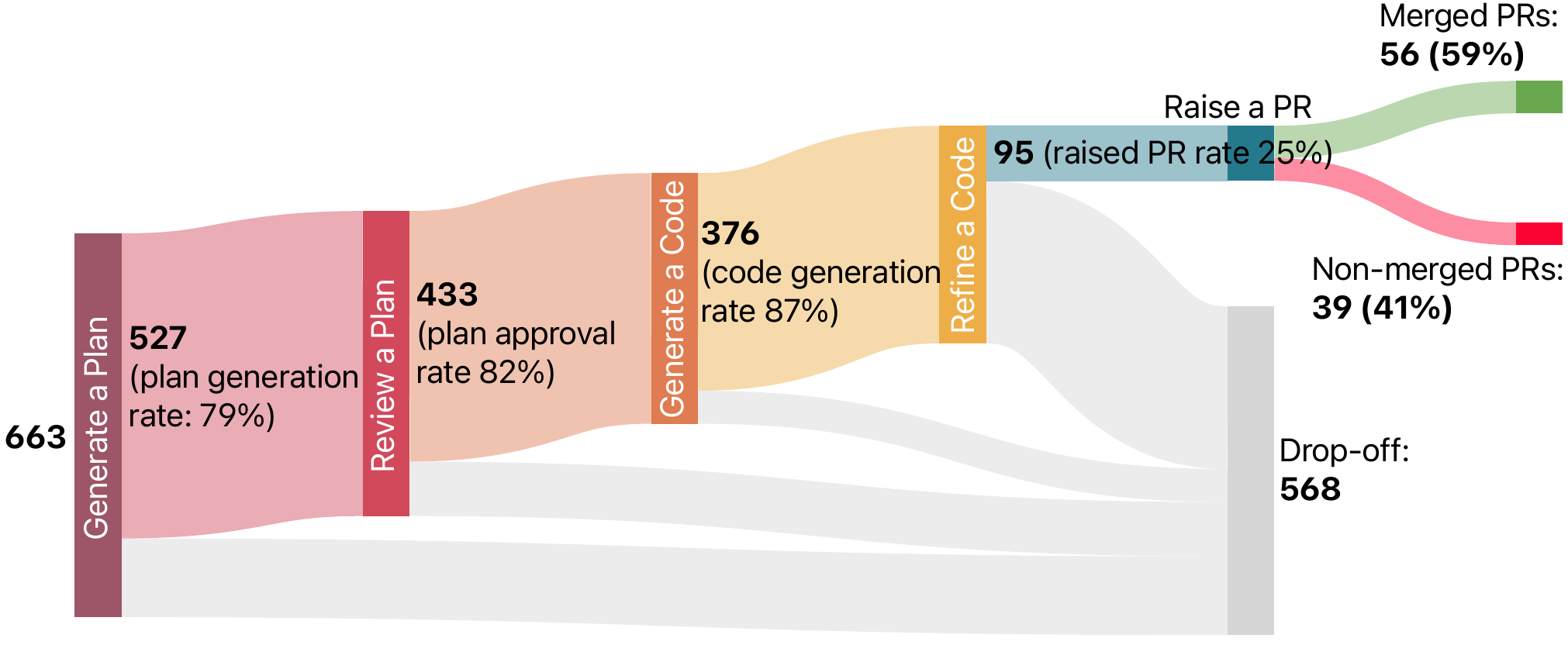}
    \caption{(RQ2) The Online Evaluation Results of \ours{}.}
    \label{fig:sankey}
\end{figure}

\subsection*{\textbf{RQ2: \rqtwo}}

The offline evaluation did not incorporate human feedback into the generation process.
Thus, this online evaluation will further assess \ours{} when incorporating human feedback from practitioners.
Figure~\ref{fig:sankey} shows the online evaluation results of \ours{}, starting from practitioners clicked \ours{} to generate a coding plan until the associated PRs were reviewed.
If a software engineer did not proceed to the next step in the \ours{} workflow, we considered the software engineer dropping out.
The usage data was recorded during the two months of deployment. 
Below, we present the key results of our online evaluation of \ours.





Figure~\ref{fig:sankey} shows that out of the 663 issues that practitioners used \ours{}, the coding plans were successfully generated for 527 issues, achieving a plan generation rate of 79\%.
Of these 527 issues, the coding plans for 433 issues were approved by practitioners to proceed to the coding stage, achieving a plan approval rate of 82\%.
These results suggest that \ours{} is generally effective at generating coding plans, with a high percentage of plans successfully passing both the generation and approval stages.


Figure \ref{fig:sankey} shows that out of 433 issues that had the coding plans approved, \ours{} successfully generated code changes for 376 issues, achieving a code generation rate of 87\%. 
Of these 376 issues, the \ours-generated code for 95 issues was then raised as PRs for review by other practitioners, achieving a raised PR rate of 25\%.
Eventually, 56 PRs out of the 95 PRs that contained the \ours-generated code were merged, achieving a merged PR rate of 59\%.

As a result, 8\% of the issues interacted with \ours{} had successfully merged \ours{}-assisted PRs into the code repositories.
These results demonstrate the success of the LLM-based software development agents in practice.

\begin{tcolorbox}[boxsep=0mm, left=2mm, right=2mm]
\textbf{Finding 2.} 
We find that during the planning stage, practitioners approved the generated coding plans for 433 out of 527 plan-generated issues, achieving a plan approval rate of 82\%. 
During the coding stage, pull requests were created for 95 out of 376 code-generated issues, achieving a raised PR rate of 25\%.
Of these, 56 pull requests were successfully merged, leading to a merged PR rate of 59\%.

\end{tcolorbox}

\subsection*{\textbf{RQ3: \rqthree}}



In September 2024, we invited 146 practitioners who had used \ours{} and completed the workflow at least once to participate in a survey. After opening the survey for three weeks, we received responses from 109 participants, achieving a response rate of 75\%.

Figure \ref{fig:demographic} shows the demographic of the participants.
Most participants were software engineers (95\%). Regarding their professional IT experience, 28\% had over 10 years of experience, 43\% had 4-9 years, 24\% had 1-3 years, and 5\% had less than one year. Additionally, 93\% of participants were familiar with AI coding assistant products, with 50\% being very familiar and 43\% somewhat familiar. Finally, 64\% had used \ours{} to generate code more than once.


Figure \ref{fig:survey_likert} presents the survey responses to our Likert-scale questions regarding the performance of \ours{} during the planning and coding stages.
Based on 109 survey responses, we generally observe low user disagreements in the planning stage.
Particularly, a total of 71\% and 62\% of participants agreed that the identified files were relevant to the issue description (Q1) and aligned with how they would have approached the task (Q2).
Considering the generated plans, 41\% and 36\% of the participants agreed that the change plans aligned with the issue description (Q3) and aligned with how they would have approached the task (Q4).
These results provide further insight into \ours{}’s strong capability to identify relevant files and plan to mimic user behaviour.

Figure \ref{fig:survey_likert} shows that 61\% of the participants agreed that the generated code could be easily understood and modified to solve the issue (Q5).
However, an adjustment is likely needed after the first generation.
Without human involvement, 33\% of the participants agreed that the generated code solved their task (Q6).
Yet, 54\% and 67\% of them disagreed that the code contains no defects and fulfilled non-functional requirements (Q7) and completely solved the task without human involvement (Q8), respectively.
These results suggest that \ours{} still faces challenges in generating the perfect code, however, the generated code is easy to modify to match participants' needs.


\begin{figure}[t]
    \centering
    \includegraphics[width=\linewidth]{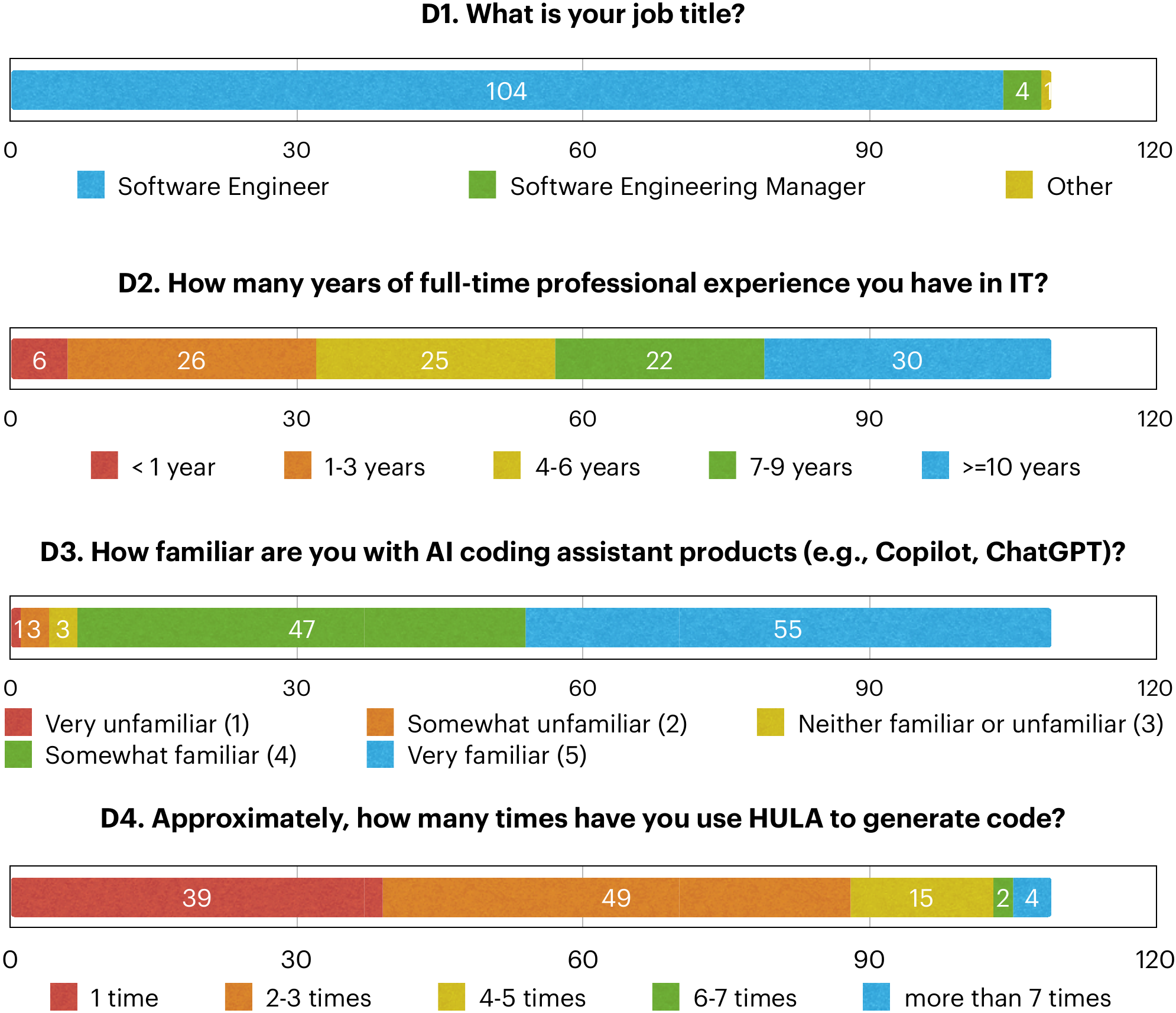}
    \caption{(RQ3) The Demographic of Participants in the Practitioners' Survey (n=109).}
    \label{fig:demographic}
\end{figure}
\begin{figure*}
    \centering
    \includegraphics[width=\textwidth]{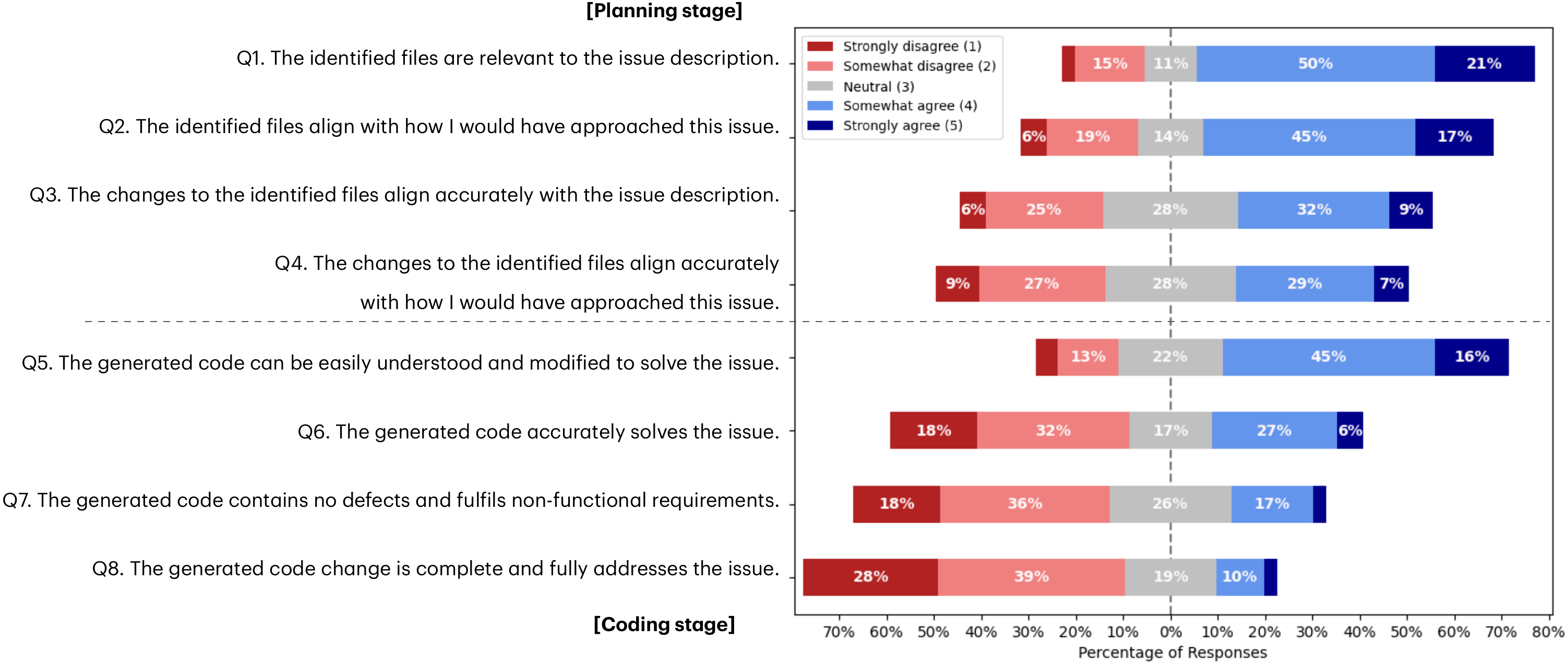}
    \caption{The Survey Responses on the Satisfaction of the Generated Plans and Code by \ours{} (n=109).}
    \label{fig:survey_likert}
\end{figure*}

Figure \ref{fig:autodev-merged} shows the analysis results of the open-ended question regarding the perceived benefits of \ours{} (Q9).
Participants highlighted several benefits, including reducing development time and effort (24 responses), resolving simple tasks (18 responses), and assisting with initiating plans and writing code (16 responses). 
Moreover, since \ours{} requires an issue description as input, a few participants (3 responses) mentioned that it promotes good documentation practices by encouraging them to provide detailed issue descriptions.

Figure \ref{fig:autodev-not-merge} presents the analysis results of the open-ended question regarding the challenges participants faced when using \ours{} (Q10). 
Consistent with the Likert-scale responses, common challenges included incorrect code functionality (25 responses) and incomplete code changes (17 responses). 
Additionally, 14 participants noted that high efforts are required to provide sufficient context for detailed issue descriptions. 
Participants also reported difficulties with the workflow, particularly with revisiting or altering the generations of previous stages (14 responses).
Other challenges included a limited awareness of external context (e.g., dependencies), incomplete code validations, difficulties in addressing large or complex tasks, and long generation times.

Figure~\ref{fig:autodev-improve} presents the analysis of the open-ended question regarding the areas of improvement for \ours{} from the participants' perspective (Q11).
Regarding the improvement in the UX/UI workflow (17 responses), participants highlighted that they prefer the flexibility to move between planning and coding stages to provide follow-up feedback (i.e., the revisiting workflow), particularly when handling partially incomplete changes.
Additionally, participants suggested automatically enhancing the input context with documentation (e.g., type definitions), best practices (e.g., design patterns), and other historical data (e.g., previous solutions) (14 responses).
Other suggestions include providing input guidance for users, enhancing code validation with selectable options, and extending capabilities to generate test cases for corresponding changes.

\begin{tcolorbox}
\textbf{Finding 3.} 
41\% of participants agreed that the generated plan by \ours{} aligned accurately with the issue description.
While \ours{} still requires human involvement to ensure the generated code completely addresses the task,  61\% of the participants agreed that the generated code was easy to read and modify, which helped reduce their initial development time and effort.
Additionally, participants acknowledged that \ours{} promotes good documentation practice, though this requires more effort to provide detailed issue descriptions.

\end{tcolorbox}

\begin{figure}[t]
    \centering
    \includegraphics[width=0.95\linewidth]{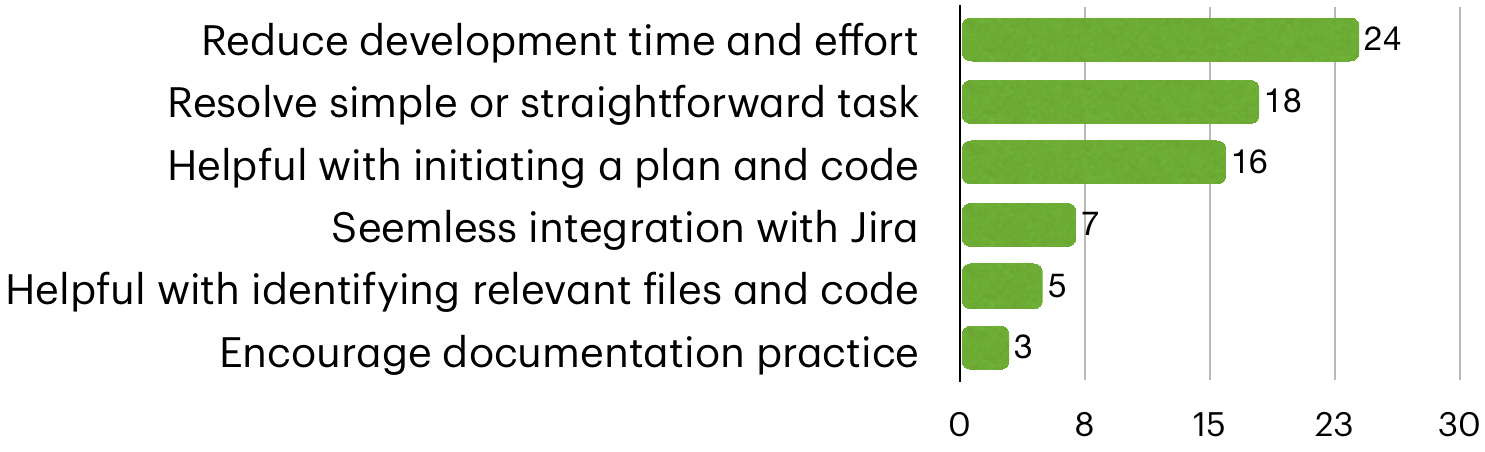}
    \caption{The Perceived Benefits of Using HULA (Q9, n=83).}
    \label{fig:autodev-merged}
    \vspace{2mm}
    \centering
    \includegraphics[width=0.97\linewidth]{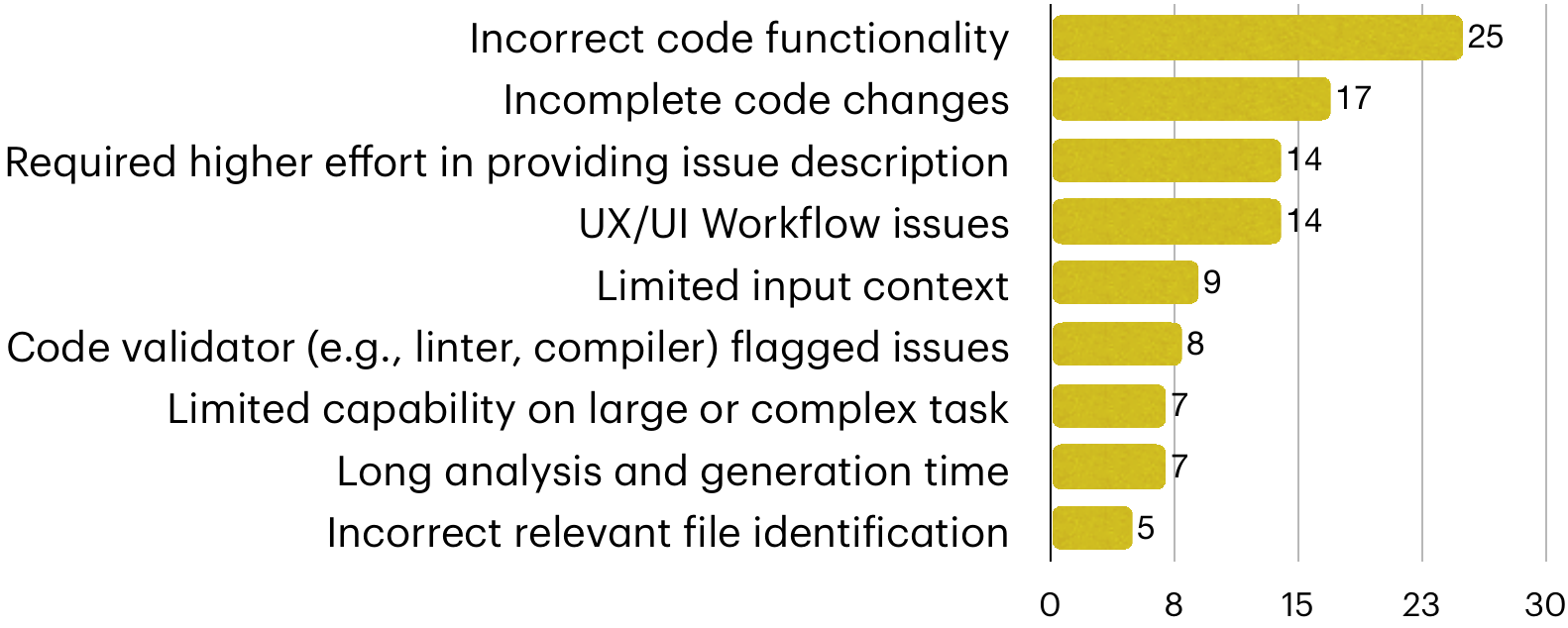}
    \caption{The Challenges Encountered when Using HULA (Q10, n=88).}
    \label{fig:autodev-not-merge}
    \vspace{2mm}
    \centering
    \includegraphics[width=0.97\linewidth]{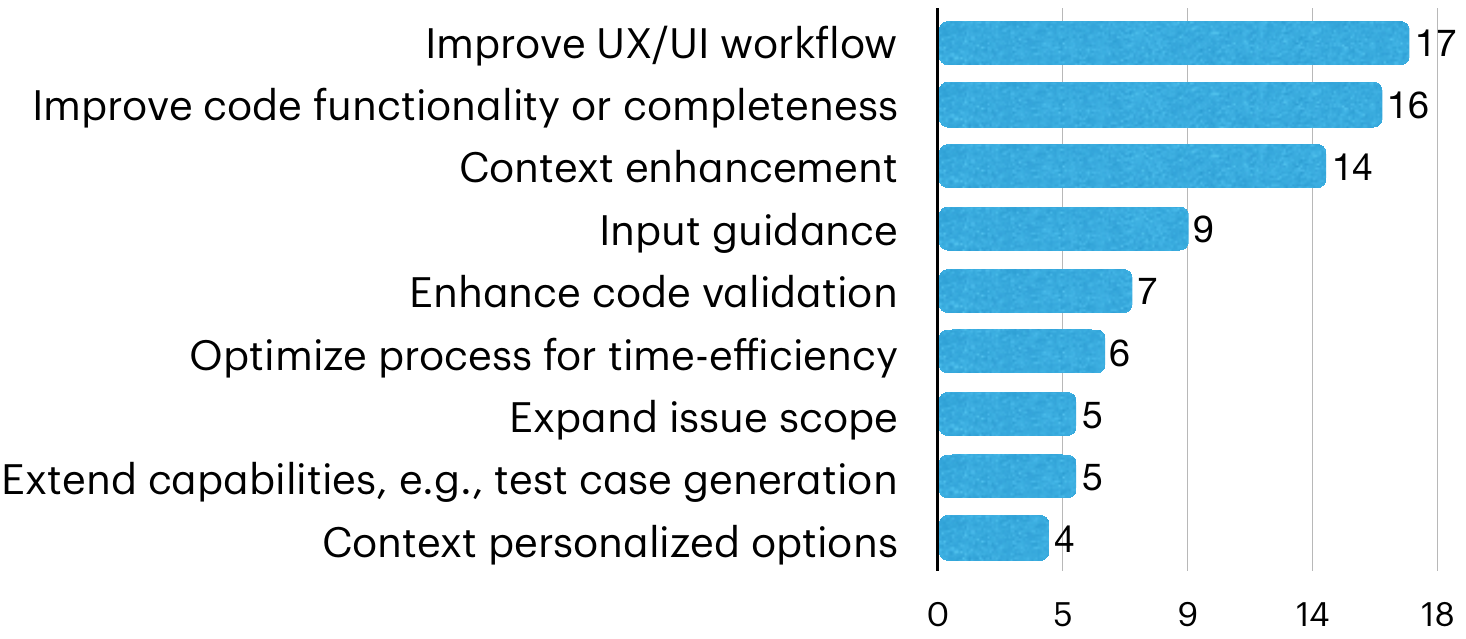}
    \caption{The Improvement Areas Suggested by Participants (Q11, n=72).}
    \label{fig:autodev-improve}
\end{figure}

\section{Lessons Learned and Future Directions}~\label{sec:future}
In this section, we discuss key lessons learned and offer suggestions for future work.


\textbf{Lesson Learned 1: The performance of an LLM-based software development agent heavily relies on a detailed input description, but what key information is needed?}
In RQ1, we found that \ours{} achieves a lower accuracy on the internal dataset when compared to the SWE-bench dataset. 
One of the possible reasons is related to the nature of how a task is written.
For the SWE-bench dataset, issues typically have a detailed description along with key information e.g., module names or code snippets.
However, for Agile-driven organizations like Atlassian, practitioners tend to collaborate closely through informal knowledge transfer (e.g., meetings, discussion threads, or chats). 
Therefore, Agile-driven JIRA issues may contain brief information like a user story or a one-liner description instead of detailed documentation (as observed in our internal dataset).

On the other hand, RQ2 indicates that practitioners' agreement with the generation is high when they can engage in the process by reviewing and enriching the issue descriptions with additional information. 
Although in RQ3, a few participants in the survey acknowledged that \ours{}'s workflow could promote good documentation practice, it requires more effort to provide detailed issue descriptions. 
To address this challenge, participants requested guidance on the needed types of input information.
Yet, little is known what key information is needed. 
Future research directions should investigate:
What types of information must be presented for LLM-based software development agents? 
How to automatically augment such information to LLM-based software development agents? 
What is the best augmentation approach for LLM-based software development agents? 

\textbf{Lesson Learned 2: Evaluating functional correctness should go beyond passing unit test cases.}
Functional correctness is the ultimate goal of any LLM-based software development agent (i.e., to generate code that is functionally correct).
During our offline evaluation, we followed the state-of-the-art approaches~\cite{hendrycks2021apps, austin2021mbpp, chen2021humaneval} to evaluate the code functionality based on the percentage of the passing unit test cases (same as SWE-bench~\cite{jimenez2023swe}).
However, our RQ3 shows that \ours{} may not always generate functionally correct code despite incorporating human and tool feedback. 

Passing unit test cases should not be the sole goal of evaluating code functionality due to the following reasons. 
First, passing unit test cases requires practitioners to prepare both test cases and the test environment (e.g., build and configuration files) to execute test cases against the code changes for a specific programming language for every pull request.
However, this approach is highly cost-intensive and intricate, particularly in our enterprise context, due to the complexity of the task and software dependencies.
Secondly, the binary nature of the test results (i.e., pass or fail) lacks the granularity required for the team to assess how closely the generated code aligns with the ground truth.
Third, test cases only show the presence of the software defects, but not their absence. 
Therefore, the generated code that passes unit test cases does not guarantee that the code is functionally correct.
To address this challenge, future research directions should explore:  
How to generate explanations or descriptions from test execution that are more beneficial to practitioners and LLM-based agents when test cases are available? 
What is an alternative approach to evaluating code functionality when test cases are not available? 
How to go beyond the traditional evaluation of functionality correctness (e.g., readability, technical debt, maintainability, etc)?

\section{Threat to validity}~\label{sec:threat}
In this section, we disclose the threats to the validity.

\textbf{Threats to construct validity.} Measuring code similarity is complex and nuanced. 
Code similarity can be measured based on various dimensions, e.g., semantic similarity, syntactic similarity, and functional similarity. 
To ensure that LLM-as-a-Judge for code similarity is highly similar to human judgement, we developed prompts that capture various concepts of code similarity.
To validate if the code similarity score is aligned with human judgement, we randomly selected 203 Jira issues.
We asked software engineers to provide a similarity score, ranging from 1 (no similarity), 2 (weak similarity), 3 (high similarity), to 4 (highly similarity).
Then, we computed the correlation efficient between the human score and LLM-as-a-Judge score. 
We observed a high correlation coefficient of 0.7, demonstrating the high degree of alignment between human judgement and LLM-as-a-judge.

\textbf{Threats to external validity.}
Our HULA framework is generally applicable to other coding platforms and other LLMs. 
However, for this paper, the findings are limited to the context of Atlassian Jira which is internally used by Atlassian software engineers and may not be generalized to other contexts. 
Therefore, the application of this framework in other coding platforms will prove fruitful.


\textbf{Threats to internal validity.}
We designed our HULA framework to be architecture-agnostic, meaning that other LLMs are applicable to this framework.
However, the experiment and findings of this paper rely on the use of GPT-4 as a backbone LLM for our \ours{} framework.
Based on the SWE-Bench Leaderboard, we find that our performance is comparable to the 6th-ranked agent as of 26 September 2024, highlighting the competitive performance of our backbone LLM. 
It is possible that other LLMs that are larger with a higher number of parameters may achieve higher performance.
Nevertheless, the key goal of this paper is not to investigate which LLMs perform best, but instead to investigate the effectiveness of the HULA framework in real-world deployment.
Therefore, the investigation of the best-performing LLMs is left for future work.

Lastly, although we may not be able to disclose all aspects of the implementation and experiment design due to industrial confidentiality reasons, we are confident that the successful development and integration of the HULA framework into Atlassian Jira is a great achievement, paving the way forward the future of AI-powered software development tools. 
In addition, our experience, findings, lessons learned, and future research directions will be beneficial for other practitioners and software engineering researchers.

\section{Conclusion}\label{sec:conclusion}

In this paper, we introduce a \underline{Hu}man-in-the-\underline{L}oop LLM-based Software Development \underline{A}gents Framework (\ours).
We design the user interface, implement the agents, and deploy the framework into Atlassian JIRA for internal uses.
Through a multi-stage evaluation of \ours{} involving real-world deployment and user surveys with Atlassian practitioners, we conclude that (1) the detail of input can highly affect the performance of \ours{}; (2) incorporating human feedback into \ours{} can enhance the input context and be beneficial in practice; (3) practitioners perceive that \ours~can help minimize the overall development time and effort but code quality remains a concern in some cases. 
While the generated plan is perceived as accurate, \ours{} still faces challenges in generating perfect code without human input.
Hence, future research is encouraged to enhance the quality of LLM-generated code~\cite{liu2024refining,she2023pitfalls}.

\section{Acknowledgments}\label{sec:ack}
This work was carried out through a collaboration between Atlassian, Monash University, and the University of Melbourne. 
The authors sincerely thank the universities and Atlassian's ADO AI team, with special appreciation to the engineering (Acrabat), data science, product, and design teams, for their invaluable support and contributions.

\section{Disclaimer}\label{sec:disclaimer}
The perspectives and conclusions presented in this paper are solely the authors' and should not be interpreted as representing the official policies or endorsements of Atlassian or any of its subsidiaries and affiliates. 
Additionally, the outcomes of this paper are independent of, and should not be construed as an assessment of, the quality of products offered by Atlassian.


\bibliographystyle{IEEEtran}
\bibliography{reference}

\end{document}